\documentclass[twocolumn,amssymb,showpacs, amsmath,nobibnotes,nofootinbib,aps, prl]{revtex4}

\begin{document}


\title{Specific heat and magnetization study on single crystals of a frustrated,
quasi one-dimensional oxide: Ca$_{3}$Co$_{2}$O$_{6}$}
\author{V. Hardy and S. Lambert}
\affiliation{Laboratoire CRISMAT, UMR 6508, Boulevard du Mar\'{e}chal Juin, 14050 Caen\\
Cedex,\\
France}
\author{M. R. Lees and D. McK. Paul}
\affiliation{Department of Physics, University of Warwick, CV4 7AL, Coventry, United\\
Kingdom}

\begin{abstract}
Specific heat and magnetization measurements have been carried out under a
range of magnetic fields on single crystals of Ca$_{3}$Co$_{2}$O$_{6}$. This
compound is composed of Ising magnetic chains that are arranged on a
triangular lattice. The intrachain and interchain couplings are
ferromagnetic and antiferromagnetic, respectively. This situation gives rise
to geometrical frustration, that bears some similarity to the classical
problem of a two-dimensional Ising triangular antiferromagnet. This paper
reports on the ordering process at low-$T$ and the possibility of
one-dimensional features at high-$T$.
\end{abstract}
\maketitle

\section{Introduction}

\noindent Ca$_{3}$Co$_{2}$O$_{6}$ belongs to a large family of compounds of
general formula A'$_{3}$ABO$_{6}$, with Ca, Sr or Ba on the A' site, while
the A and B sites can be occupied by a great variety of cations including
many transition metal elements.\cite{ST01} The rhombohedral structure of
these compounds consists of $\left[ \text{ABO}_{6}\right] _{\infty }$%
infinite chains running along the $c$-axis of the corresponding hexagonal
cell, with the A' cations located in between them (see Fig. 1). These chains
are made of alternating, face-sharing AO$_{6}$ trigonal prisms and BO$_{6}$
octahedra. Each chain is surrounded by six equally spaced chains forming a
triangular lattice in the $ab$ plane. The intrachain A-B separation is quite
small, whereas the interchain distance is approximately twice that distance
(for example about 2.6 and 5.2 \AA , respectively, in the case of Ca$_{3}$Co$%
_{2}$O$_{6}$).\cite{FJ96} Reinforcing this geometrical anisotropy, the
oxygen atoms along the chains can efficiently mediate the intrachain
coupling, whereas the A' cations play no role in the interchain coupling.
Therefore, these compounds are expected to display pronounced
one-dimensional (1D) magnetic character.

Amongst this family, Ca$_{3}$Co$_{2}$O$_{6}$ has attracted special interest
in recent years due to its very peculiar magnetic properties.\cite
{AA97,KA97a,KA97b,MA00} According to the most recent studies, \cite
{KA97b,MA00} both the cobalt ions in this compound are trivalent, but as
they are subject to different crystalline-electric-fields (CEF), the Co ions
are in different spin states. As pointed out by Aasland {\it et al.},\cite
{AA97} the CEF-induced splitting of the 3d orbitals produces a gap between a
lower triplet and an upper doublet that is larger in an octahedral
environment than in a prismatic one. As a consequence, the 3d$^{6}$
configuration of Co$^{3+}$ can produce a high-spin state ($S=2$) on the
prismatic site and a low-spin state ($S=0$) on the octahedral site. Although
indisputable evidence is still absent, this scenario is the most consistent
with the neutron diffraction data and the magnetic measurements that are
available at the present time.\cite{AA97,KA97a,KA97b,MA00} At high
temperatures, it must be borne in mind that the situation may be complicated
by a progressive transition from the low-spin state to the intermediate-spin
state ($S=1$) and/or the high-spin state.\cite{MA00}

The previous studies on Ca$_{3}$Co$_{2}$O$_{6}$ have established several
basic features about the Co-Co interactions in this compound: (i) a strong
Ising-character,\cite{KA97c} with the spins oriented along the chain axis;
(ii) a ferromagnetic intrachain coupling; (iii) an antiferromagnetic
(nearest-neighbor) interchain coupling. Such features, combined with the
triangular arrangement of the chains on the $ab$ plane, give rise to a
prototypical situation of geometrical frustration. Ca$_{3}$Co$_{2}$O$_{6}$
is also reported to exhibit peculiar magnetic behaviour, especially in $M(H)$
curves.\cite{AA97,KA97a,KA97b,MA00} For instance, a sequence of irreversible
metamagnetic transitions take place at very low $T$ (%
\mbox{$<$}%
5 K), in the so-called Frozen Spin state, while there is a sharp
ferrimagnetic-to-ferromagnetic transition at intermediate temperatures
around 10 K. Complex phase diagrams have been proposed.\cite{KA97a,MA00}
Some features of these phase diagrams are still the subject of controversy.
One of the most debated issues deals with the ordering process under zero or
low field. Kageyama {\it et al.}\cite{KA97a} have proposed a scenario
similar to the one developed by Mekata {\it et al.}\cite{ME77,ME78} for
CsCoCl$_{3}$ (antiferromagnetic Ising chains, antiferromagnetically coupled
on a triangular lattice).\cite{CO97} Upon cooling, this model predicts an
initial transition to a Partially Disordered Antiferromagnetic (PDA) state,
in which two thirds of the chains are antiferromagnetically coupled while
the last third remains incoherent,\cite{rqmartin} followed by a second
transition to a ferrimagnetic state at lower $T$. Whether this scenario is
applicable to the case of Ca$_{3}$Co$_{2}$O$_{6}$ has recently been
questioned.\cite{MA00,Ni01} To date no specific heat measurements have been
performed on this compound, although such thermal data would be valuable in
understanding the physics of this material.

In this paper, we report on specific heat measurements under an applied
magnetic field, carried out on Ca$_{3}$Co$_{2}$O$_{6}$ single crystals. The
magnetic field has allowed us to investigate the ordering process in the
various magnetic states adopted by this material: disordered
antiferromagnetic, ferrimagnetic and ferromagnetic. Such a study had to be
carried out on single crystals due to the highly anisotropic properties of
this compound.\cite{KA97b,MA00} This paper is organized as follows. Section
II is devoted to experimental details, and in particular details of the
single-crystal alignment procedure and the subtraction of the lattice
contribution to the specific heat. Section III reports the specific heat
data under zero field and large magnetic fields, along with the
corresponding magnetization measurements. The results are discussed in
Section IV and our conclusions are presented in Section V.

\section{Experimental details}

\bigskip Single crystals were grown using a flux method described
previously. \cite{MA00} This method yielded needle-like crystals - similar
to those used for magnetic measurements in Ref. 6 -along with other
specimens that have an aspect ratio that is more suitable for specific heat
measurements. These latter crystals have the shape of short hexagonal rods
terminating in three diamondlike faces at each end. Their small size ( $%
0.7\times 0.7\times 1$ mm$^{3}$) makes it difficult to get reliable specific
heat data on only one crystal. We have used a set of four crystals with a
total mass of 10.64 mg. Magnetic measurements carried out on each crystal
verified the good homogeneity of this assembly.

All magnetic and specific heat measurements reported in this paper were
recorded with the magnetic field along the $c$ axis of the crystals. The
crystals were aligned using the following procedure: (i) at 300 K, the four
crystals were put into a small amount of Apiezon N grease (contained in a
gelatine capsule, or on the measuring platform, for magnetization or
specific heat, respectively); (ii) a magnetic field of 5 T is then applied.
Owing to the substantial magnetic anisotropy persisting up to room
temperature in this compound, the samples experience a torque which tends to
align their $c$-axis along the direction of the applied field. We checked
that 5 T is large enough to rotate our crystals in Apiezon N at 300 K, a
temperature at which this grease has a low viscosity; (iii) the samples are
then slowly cooled under 5 T down to 150 K. This ensures the field alignment
is maintained through the glass transition of Apiezon N\ grease that occurs
around 230 K;\cite{KR72} (iv) at 150 K, the grease is solid enough to hold
the samples in position without a magnetic field, while the temperature is
still above the magnetic transitions of the compound. The field is reduced
to zero, and the cooling is resumed to prepare the samples in a Zero Field
Cooled (ZFC) state. The quality of the alignment obtained by this method was
demonstrated by the sharpness of the ferrimagnetic-to-ferromagnetic
transition and the magnetization values at the associated plateaus.

The specific heat measurements were carried out by a two-tau relaxation
method (PPMS, Quantum Design) with magnetic fields up to 5 T and
temperatures down to 2 K. A background signal (platform and grease) was
recorded versus temperature under each field investigated. Magnetization
measurements were carried out on the same set of crystals, by means of a
Superconducting Quantum Interference Device magnetometer (MPMS, Quantum
Design), with magnetic fields up to 5 T and temperatures down to 2 K.

In order to subtract the lattice contribution to heat capacity, a
non-magnetic isostructural compound was synthesized. A\ ceramic sample of Ca$%
_{4}$PtO$_{6}$ was prepared from a stoichiometric mixture of CaO and PtO$%
_{2} $. These starting materials were pelleterized in the form of bars and
heated in a evacuated silica ampoule at 800$^{\circ }C$ for 12 h. X-ray
diffraction verified the single phase of the Ca$_{4}$PtO$_{6}$ compound, and
confirmed that this material has the same rhombohedral structure as Ca$_{3}$%
Co$_{2}$O$_{6}$. In Ca$_{4}$PtO$_{6}$, the Ca$^{2+}$ cations occupy the
trigonal prisms, while the Pt$^{4+}$(3d$^{6}$) on the octahedral sites are
supposed to be in a low-spin state ($S=0$) due to a large CEF-induced gap.
To check this point, magnetic measurements were performed between 2 and 300
K, leading to a susceptibility of the form $\chi =\chi _{0}+C\,/\,T$, with $%
C=0.00151$ emu K mol$^{-1}$. This Curie constant is about 0.05\% of the
value expected for high-spin Pt$^{4+}$ ($S=2$), and 0.15\% of the value
expected for intermediate-spin Pt$^{4+}$ ($S=1$). This result ensures that
the specific-heat data of Ca$_{4}$PtO$_{6}$ (up to 300 K) does not contain a
significant magnetic contribution from the electronic spins. Furthermore,
the data down to 2 K for both compounds reveals no contribution from a
Schottky term related to nuclear level splitting. Finally, one can also
safely discard the presence of a significant electronic term in these
compounds owing to their poor electrical conductivity.\cite{MA00,RA02} One
can thus consider that the specific heat of Ca$_{3}$Co$_{2}$O$_{6}$, in the
range 2-300 K, is made up of two terms: $C(T)=C_{L}(T)+C_{M}(T)$, where $%
C_{L}$ is the lattice contribution related to phonon excitations, while $%
C_{M}$ is the magnetic term that we are interested in.

The contribution $C_{L}(T)$ of Ca$_{3}$Co$_{2}$O$_{6}$ can be derived from
the total specific heat $C^{\prime }(T)$ of Ca$_{4}$PtO$_{6}$, but the
difference in molecular weight between the two compounds must be taken into
account. There is no generally accepted method to do this, and some
approximations have to be made. Within the framework of the Debye approach,
the lattice contribution is determined only by the ratio $\theta /T$, where $%
\theta $ is the Debye temperature, which depends on the structure and the
mass of the constituent atoms.\cite{GOPAL} Accordingly, the correction of
mass amounts to the determination of the ratio $r=\theta \,/\,\theta
^{\prime }$, where $\theta $ and $\,\theta ^{\prime }$ are the Debye
temperatures of Ca$_{3}$Co$_{2}$O$_{6}$ and Ca$_{4}$PtO$_{6}$, respectively.
Since $\theta $ in the basic monoatomic model is inversely proportional to
the square root of the mass, one can get a first estimate by using the
molecular weights, leading to $r\simeq \sqrt{M(Ca_{4}PtO_{6})\,/%
\,M(Ca_{3}Co_{2}O_{6})}=1.16$. A more sophisticated method based on the
principle of corresponding states has been developed by Stout and Catalano. 
\cite{STOUT55} This technique is based on the comparison of the specific
heat data at high temperatures for the magnetic and the reference compounds.
It has the great advantage of accounting for a temperature dependence in the
parameter $r$, but, on the other hand, it requires an extrapolation of $r$
in the low $T$\ range (below about 190 K in our case) which introduces a
significant uncertainty. Following the extrapolations proposed in Ref. 16,
we found $r\sim 1.15$ at low $T$. In the present study, we also have the
opportunity to derive an experimental estimate of $r$ by analysing the
low-temperature dependence of the specific heat data. Indeed, the lattice
contribution is reduced to a term $\beta _{3}T^{3}$ at low temperature, with 
$\beta _{3}\propto \left( 1/\theta \right) ^{3}$. In a $C/T$-vs-$T^{2}$
plot, both the curves of Ca$_{3}$Co$_{2}$O$_{6}$ and Ca$_{4}$PtO$_{6}$
exhibit a linear regime at low $T$ from which one can extract $\beta _{3}$
values leading to $r\simeq \left[ \beta _{3}(Ca_{4}PtO_{6})\,/\,\beta
_{3}(Ca_{3}Co_{2}O_{6})\right] ^{1/3}=1.13$.

In the analysis presented hereafter, we chose to retain this latter value,
which appeared to us as the best compromise between simplicity and
reliability. However, we will take the uncertainty on the $r$ value into
account when discussing the comparison of $C_{M}(T)$ with theoretical
predictions (Sec. IV-B). Since one can write $C_{L}(T)=C^{\prime }(T\,/\,r)$
within the framework of the Debye model, the lattice contribution of Ca$_{3}$%
Co$_{2}$O$_{6}$ has been derived by multiplying the temperature values of
the total specific heat data $C^{\prime }(T)$ of Ca$_{4}$PtO$_{6}$ by\ $%
r=1.13$.

\section{Results}

Figure 2 shows the total specific heat versus temperature for Ca$_{3}$Co$%
_{2} $O$_{6}$ measured under zero-field, along with the estimated lattice
contribution obtained by using the procedure described above. The main
feature is a prominent peak in the raw data around 25 K, indicative of a
long-range ordering (LRO). It can also be observed that the total specific
heat $C(T)$ tends to merge with the lattice contribution $C_{L}(T)$ at high $%
T$. The inset displays an enlargement of the low $T$ portion of the data as
a $C/T$-vs-$T^{2}$ plot. Three separate runs are shown together, which
illustrate the good reproducibility of the $C(T)$ data. This plot also shows
that the lattice term has a simple cubic temperature dependence ($%
C_{L}=\beta _{3}T^{3})$, as expected for such contributions. We found $\beta
_{3}\simeq 3$ $10^{-4}$ J K$^{-4}$ mol$^{-1}$, leading to a Debye
temperature $\theta \simeq 415$ K. Beside this lattice term, the specific
heat of Ca$_{3}$Co$_{2}$O$_{6}$ exhibits a linear contribution $\gamma T$ ($%
\gamma \simeq 10$ mJ K$^{-2}$ mol$^{-1})$, and one can see the development
of magnetic excitations above $\sim 7$ K.

Figure 3 shows the magnetic part of the specific heat, $%
C_{M}(T)=C(T)-C_{L}(T)$, under zero-field. Beyond the sharp peak around 25
K, \cite{rq0} there is a broad bump at higher temperatures. Because of
increased scatter in the data at high $T$, which is further enhanced by the
subtraction procedure, the curve has been cut above 200 K. The inset shows
the temperature dependence of the magnetic entropy $S_{M}(T)=\int_{0}^{T}%
\left( C_{M}\,/\,T\right) \,dT$. The peak in $C_{M}(T)$ at 25 K yields a
significant change in the rate of entropy variation around this temperature.
It should be noted that the entropy at the LRO transition, $S_{M}(25$ K$)$
is equal to 18 \% of the total magnetic entropy. This small value is
consistent with the expected behaviour in a 1D system.\cite{REV} One can
also note that $S_{M}$ at high $T$ tends to saturate just below the expected
value for one cobalt of spin 2 per formula unit ($R\,\ln 5)$.

Before showing the specific heat results under magnetic field, let us
consider the corresponding magnetization data. Figure 4 shows an enlargement
of $dc$-susceptibility curves for an applied field of 0.1 T, 2 T and 5 T,
and focusses on the region around the transition at 25 K. In both 2 T and 5
T, one observes a small difference between the Zero Field Cooled (ZFC) and
Field Cooling (FC) curves at very low $T$ ($T$ 
\mbox{$<$}%
5 K), a behaviour which is consistent with the large hysteresis displayed by
the $M(H)$ curves in the same temperature range.\cite{KA97a,KA97b,MA00}
Above this irreversible regime, the low $T$ plateaus of the curves under 2 T
and 5 T correspond to magnetizations equal to $M_{S}\,/\,3$ and $M_{S}$,
respectively, where $M_{S}$ is the full spin polarization.\cite{rq1} This is
consistent with the ferrimagnetic and ferromagnetic natures of the low $T$
states in 2 T and 5 T, respectively, that have been reported previously. 
\cite{AA97,KA97a,KA97b,MA00} Let us now turn to the corresponding specific
heat data recorded in 0 T, 2 T and 5 T after ZFC. Even below 5 K, no
significant hysteresis could be detected between the ZFC and the FC curves
in the $C(T)$ data. Furthermore, we checked that there is no difference
between the $C(T)$ curves recorded under 0 T or 0.1 T. Figure 5 shows the $%
C_{M}(T)$ curves under field which were derived by subtracting $C_{L}(T)$
from the raw data.

The 2 T data exhibit a very large, lambda-like peak at 25 K. It must be
emphasized that the height of this peak is much larger than that found under
0 T. In 5 T, there is only a smooth bump around 25 K. This is consistent
with the expected behaviour for a long-range ferromagnetic transition in a
large field. In the low $T$ regime, one observes that the 2 T curve lies
well below the data collected with no field applied. One can get a better
insight into this low $T$ range by considering the temperature dependence of
the magnetic entropy as displayed in Fig. 6(a). The curves for 0 and 2 T are
close to each other below about $7$ K and above $T_{N}$. In the intermediate
range, the entropy in 0 T is much larger than in 2 T.

Figure 6(b) shows the ZFC susceptibility curves in the same temperature
range, that were recorded under 0.1 T with a waiting time $t_{W}$ after each
temperature stabilization equal to either 5 seconds or 600 seconds. Although 
$\chi (T)$ starts increasing abruptly at $T_{N}$ (see also Fig. 4), the
maximum slope in the ZFC curves is actually found at $\sim 18$ K. In between 
$18$ K and $8$ K, the ZFC\ curve exhibits a pronounced relaxation effect.
Below $8$ K, $\chi (T)$ becomes time-independent and is roughly constant
down to 2 K.

\section{Discussion}

\subsection{The ordering process under low field}

In the literature on Ca$_{3}$Co$_{2}$O$_{6}$, the existence of two critical
temperatures at $T_{C_{1}}$ ($\simeq 25$ K) and $T_{C_{2}}$ ($\simeq 10$ K)
has often been invoked: the first has been associated with a ferromagnetic
intrachain transition, while the second is identified with an
antiferromagnetic transition between the ferromagnetic chains. One of the
main pieces of information provided directly by our specific heat
investigations is the sharp peak observed at 25 K in the $C(T)$ data under
zero-field. This feature clearly demonstrates the occurrence of a collective
magnetic transition. Since it is well known that such long-range ordering
(LRO) cannot take place in one dimension, this feature cannot be ascribed to
a ferromagnetic transition along the chains. The scenario described above is
thus inconsistent with the specific heat data. We propose that there is a
collective transition at $T_{N}\simeq 25$ K corresponding to an
antiferromagnetic coupling between the chains. Note this is consistent with
the emergence of antiferromagnetic Bragg peaks around 25 K in neutron
diffraction data.\cite{AA97} However, the entropy jump associated with this
transition shows that the antiferromagnetic ordering is far from being
complete. At the same time, the development of a long-range interchain
ordering, even if it is partial, can reduce the magnetic fluctuations along
the chains. This could explain why the antiferromagnetic ordering at $T_{N}$
is accompanied by a sudden increase of the susceptibility.

The existence of a LRO transition at a rather high temperature in Ca$_{3}$Co$%
_{2}$O$_{6}$ emphasizes a clear departure from the pure Ising triangular
antiferromagnet which has a totally frustrated ground state.\cite
{WA50,ME74,SA84} One of the most likely origins for such a lifting of
degeneracy is the influence of next-nearest interchain interactions.\cite
{rq2} Some studies of Ca$_{3}$Co$_{2}$O$_{6}$ proposed a scenario involving
a PDA (Partially Disordered Antiferromagnetic) state,\cite{KA97a,KA97b}
which implies AF nearest-neighbor coupling and F next-nearest--neighbours
coupling. In this model, the decrease of temperature first induces a
transition at $T_{N1}$ to a PDA state, followed at $T_{N2}$ by a transition
to a ferrimagnetic state.\cite{ME77,ME78} This process is supposed to give
rise to a smooth peak in specific heat at $T_{N\,1}$ and a large one at $%
T_{N\,2}$. This is in contrast with our experimental results in Ca$_{3}$Co$%
_{2}$O$_{6}$, since we observed a sharp peak at the first transition, while
no additional peak exists at lower temperatures. It should be noted,
however, that even in the ABX$_{3}$ compounds (A=Cs, Rb, and B=Co) for which
this model was developed, the specific heat data do not show clear support
for the theoretical expectations.\cite{YE75,AM90,WA94} It was argued that
the transition to ferrimagnetism might be prevented by the establishment of
a glassy phase at low temperatures.\cite{TA95} In Ca$_{3}$Co$_{2}$O$_{6}$,
this latter phase could be the Frozen Spin state which was found in
magnetization studies.\cite{KA97a,KA97b,MA00} Our specific heat measurements
show that the transition to the Frozen Spin state ($T_{FS}\simeq 7$ K) is
associated with a noticeable change of regime in a $C/T$-vs-$T^{2}$ plot
(see inset of Fig. 2). In the low $T$ regime, the specific heat exhibits a
linear term with $\gamma \simeq 10$ mJ K$^{-2}$ mol$^{-1}$. Such a feature
is usually found in spin glasses\cite{MYDOSH} and in other disordered
magnetic systems such as mixed-valent manganites.\cite{GH99,SM00} We can
thus speculate that this linear contribution reflects the disorder that
remains in the chain system of Ca$_{3}$Co$_{2}$O$_{6}$ at low $T$.

As for the shape of the peak at $T_{N\,1}$, it was found that it could vary
considerably with the level of frustration.\cite{TA95} Depending on the
signs and respective values of the first, second and third neighbor
interactions in a 2D Ising triangular antiferromagnet, Monte Carlo
simulations showed that a large variety of $C(T)$ curves can be expected,
including a single large peak at $T_{N\,1}.$ Our observations are thus
consistent with such a generalized PDA scenario, while they seem to rule out
the standard picture proposed earlier.\cite{KA97a,KA97b}

The specific heat measurements provide us with information about the entropy
variation in between $T_{FS}$ and $T_{N}$. In an applied field of 2 T, Ca$%
_{3}$Co$_{2}$O$_{6}$ undergoes a transition from a disordered state (system
of uncoupled chains) to a ferrimagnetic state. The specific heat data
exhibit a very large, sharp peak at $T_{N}$, which is much more pronounced
than under zero-field. The magnetic entropies calculated at $T_{N}$,
however, are found to be the same under both fields. Therefore, the magnetic
field only affects the way the entropy evolves below $T_{N}$. In an applied
field of 2 T, the magnetic phase at low temperature is well ordered
(ferrimagnetic state), so there is a sudden large release of entropy just
around $T_{N}$. In contrast, the transition under zero-field is more
progressive. Although there is a noticeable change of entropy at $T_{N}$, it
is much smaller than under 2 T, and the entropy is progressively reduced as
the temperature is decreased further. The difference between the $S(T)$
curves under 0 T and 2 T can give direct insight into the large magnetic
disorder that persists far below $T_{N}$ under zero field. Upon cooling
below $T_{N}$, the slow decrease of entropy in zero-field is interrupted by
the spin freezing at $T_{FS}\simeq 7$ K. This is also the temperature at
which the $S(T)$ curves at 0 T and 2 T\ are found to merge onto each other.

Another aspect of this ordering process is revealed by magnetization data at
low field, which showed that a pronounced relaxation effect takes place
below $T_{N}$. This effect starts above $T_{FS}$, which is consistent with
the fact that one cannot expect significant effects over short time scales
(i.e. 600 seconds) in a frozen spin state. This large relaxation effect
[Fig. 6(b)] indicates that the magnetic disorder revealed by the specific
heat measurements [Fig. 6(a)] evolves with time. This behaviour is probably
related to the frustration and the slow dynamics associated with spin
reversals in ferromagnetic chains.\cite{KA97b,MA00}

\subsection{The broad maximum at high temperature}

Another striking feature of the magnetic contribution to the specific heat
is the existence of a broad maximum at high temperatures, well above $T_{N}$%
. Actually, such a result is theoretically expected in a 1D magnetic
compound, owing to the development of short-range ordering along the chains. 
\cite{REV} From an experimental point of view however, there are not so many
reports on such features, in particular for Ising systems. To our knowledge,
the only clear report of such a broad maximum with Ising ferromagnetic
chains dealt with CoCl$_{2}\bullet 2$NC$_{5}$H$_{5}$.\cite{TA71} The
temperature of the maximum ($T_{\max })$ in $C_{M}(T)$ as well as its height
($C_{\max })$ are predicted to increase with the spin value.\cite{REV} This
could explain why this broad maximum is so visible in our case while it is
generally hidden by the LRO transition in Ising systems with smaller spin
values. According to theoretical predictions for a Ising system with $S=2$,
one should expect $T_{\max }\simeq 3.72\;J$ and $C_{\max }\simeq 1.4\;R,$
where $J$ and $R$ are the intrachain coupling and the gas constant,
respectively.\cite{REV,OB68} In our case, $C_{\max }$ is found to be close
to $8$ rather than $1.4\;R\sim 12$ J K$^{-1}$mol$^{-1}$, but it should be
noted that $C_{\max }$ is quite sensitive to the value of the $r$ parameter
used in the subtraction of the lattice contribution (e.g. one finds $C_{\max
}\sim $ 11 J K$^{-1}$mol$^{-1}$ for $r\sim 1.16)$. In contrast, $T_{\max }$
is found to be essentially independent of the $r$ value in the range $%
1.13-1.16$. This characteristic temperature leads to $J\sim 22$ K, while
Kageyama {\it et al.} reported $J\sim 15$ K from an analysis of the
susceptibility at high $T$.\cite{KA97b} Taking into account that this latter
analysis may also contain some uncertainties, one can consider that there is
a reasonable overall agreement between the experimental $C_{M}(T)$ data and
the theoretical expectations for short-range excitations in an $S=2$ Ising
chain.

It must be emphasized that a totally different phenomenon may contribute to
the existence of a broad peak in $C_{M}(T)$. The cobalt ions in the
octahedral sites that are in low-spin (LS) state at low $T$ could change
their spin state to high or intermediate spin states (HS or IS) as $T$\ is
increased. Such a phenomenon is currently the source of intensive debate\cite
{LaCo,YA96,ZO02,NO02} in the case of LaCoO$_{3}$, which also contains LS Co$%
^{3+}$ on octahedral sites at low $T$. Rather than a true spin state
transition, an increase in temperature can induce a progressive thermal
population of the IS and/or HS levels. Such a process will have a pronounced
signature in $C_{M}(T)$, in the form of a Schottky anomaly.\cite{GOPAL}
Therefore, a spin-state crossover, if it occurs, could contribute to the
bump seen in specific heat at high $T$.\cite{rq3} Accordingly, it is worth
noting that the existence of such an additional phenomenon could affect to
some extent the analysis of the bump in $C_{M}(T)$ in terms of pure
short-range spin ordering. The interplay between the short-range spin
ordering and a possible spin-state crossover must be very complex, and at
the present time, it seems difficult to separate the two effects simply on
the basis of specific heat and/or magnetization data. It is clear that more
direct investigations of the Co spin states in Ca$_{3}$Co$_{2}$O$_{6}$ are
required to proceed further with this analysis.

\section{Conclusion}

Our specific heat investigations in Ca$_{3}$Co$_{2}$O$_{6}$ have revealed
several features that complement the previous magnetization studies. We have
shown that there is a peak at $T_{N}\simeq 25$ K in $C_{M}(T)$ under an
applied field of 0 T and 2 T. This feature clearly demonstrates the
occurrence of a long-range ordering associated with the antiferromagnetic
interchain coupling. In 5 T, there is no longer a peak in the data, as
expected for a ferromagnetic transition in a large magnetic field. In an
applied field of 2 T, the ordered state is ferrimagnetic and the peak at $%
T_{N}$ is found to be very pronounced. Under zero field, the peak at $T_{N}$
is significantly reduced; the magnetic entropy displays a smooth, continuous
evolution versus temperature below $T_{N}$, that is interrupted when
entering the Frozen Spin state around $T_{FS}\simeq 7$ K. One observes a
pronounced crossover in the temperature dependence of $C_{M}(T)$ around $%
T_{FS}$. Below $T_{FS}$, the specific heat has a linear term ($\gamma T$)
that can be associated with frozen magnetic disorder. It must be emphasized
that no additional peak was detected in $C_{M}(T)$ below $T_{N}$, which
appears to rule out a ferrimagnetic transition as proposed by some PDA
scenarios.\cite{ME77,ME78}

Magnetization measurements revealed a noticeable time dependence in the
intermediate temperature range between $T_{FS}$ and $T_{N}$. The combination
of our specific heat and magnetization results demonstrates that the
magnetic state below $T_{N}$ under zero field is still highly disordered and
evolves continuously with both temperature and time. This feature probably
derives from a combination of geometric frustration and the slow spin
dynamics of ferromagnetic chains.

The magnetic heat capacity exhibits a broad maximum at high $T$ which can be
related to one-dimensional short-range ordering along the ferromagnetic
chains. The fact that only a small fraction ($\simeq $ 0.18) of the total
magnetic entropy $[R$ $ln(2S+1)]$ is released up to $T_{N}$ is consistent
with the expectations for a 1D system. It is worth emphasizing that Ca$_{3}$%
Co$_{2}$O$_{6}$ and a few other compounds of the same family\cite{other}
seem to be the only model systems for the case of Ising ferromagnetic
chains, antiferromagnetically coupled on a triangular lattice.

\bigskip The authors thank C. Martin for the synthesis of the $Ca_{4}PtO_{6}$
compound. This work is supported by a EPSRC Fellowship grant (GR/R94299/01$)$
to one of the authors (V. H. ). The authors also acknowledge financial
support from the CNRS under a CNRS / Royal Society exchange program (n$%
^{\circ }$ 13396).

\section{Figure captions}

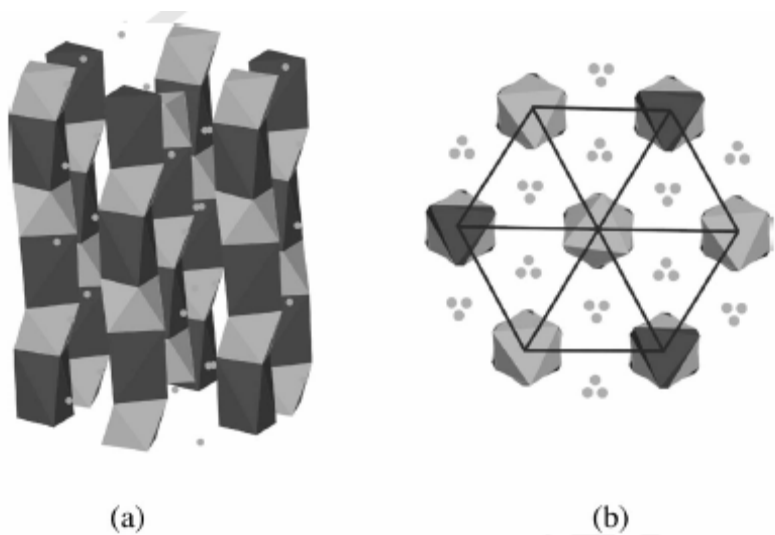
Fig. 1: Schematic drawings of the structure of A'$_{3}$ABO$_{6}$-type
compounds. The light and dark polyhedra represent AO$_{6}$ trigonal prisms
and BO$_{6}$ octahedra, respectively. The shaded circles denote A' atoms.
(a) Perspective view showing the $\left[ \text{ABO}_{6}\right] _{\infty }$%
chains running along the hexagonal $c$-axis. (b) Projection along the
hexagonal $c$-axis. Solid lines emphasize the triangular arrangement of the
chains in the $ab$ plane.

Fig. 2: Total specific heat ($C$) and lattice contribution ($C_{L}$) in Ca$%
_{3}$Co$_{2}$O$_{6}$ under zero-field. The inset shows the low $T$\ range in
a $C/T$-vs-$T^{2}$ plot, with three separate runs for $C(T)$.

Fig. 3: Temperature dependence of the magnetic specific heat in zero field.
The inset shows the calculated magnetic entropy as a function of
temperature, along with the expected total value R$\ln 5$ (dashed line).

Fig. 4: Temperature dependence of the $dc$-susceptibility under 0.1 T
(circles), 2 T\ (squares) and 5 T (diamonds), in the ZFC (open symbols) and
FC (closed symbols) modes. The field is applied along the $c$ axis
(direction of the chains and of the spins).

Fig. 5: Temperature dependence of the magnetic specific heat in 0 T
(circles), 2 T\ (squares) and 5 T (diamonds). The field is applied along the 
$c$ axis (direction of the chains and of the spins).

Fig. 6: (a) Temperature dependence of the magnetic entropy in 0 T (circles),
2 T\ (squares) and 5 T (diamonds). The field is applied along the $c$ axis
(direction of the chains and of the spins); (b) Temperature dependence of
the ZFC susceptibility under 0.1 T (applied along the $c$ axis) for two
values of waiting time after each temperature stabilization: 5 seconds
(circles) and 600 seconds (triangles).


\begin{thebibliography}{99}
\bibitem{rq0}  The exact temperature of the peak was found to be $24.7\pm
0.1 $ K
\bibitem{rq1}  From the high field plateau of ferrimagnetic-to-ferromagnetic
transitions around 10 K, we found $M_{S}\simeq $ 4.6 $\mu _{B}$ / f.u.. This
value, that is slightly larger than those previously reported in the
literature, is consistent with one cobalt of spin equal to 2 per formula
unit, with a Lande factor close to 2.3.
\bibitem{rq2}  Owing to the 3D character of the present situation, such
next-nearest interactions could originate from the nearest-neighbouring
chains but involving magnetic cobalt shifted along the $c$ axis.
\bibitem{rq3}  Note that the gap values reported in the literature for LaCoO$%
_{3}$\cite{YA96,ZO02,NO02} can give rise to a Schottky anomaly in a
temperature range consistent with the broad peak in $C_{M}(T)$ observed in
our data.
\end{thebibliography}
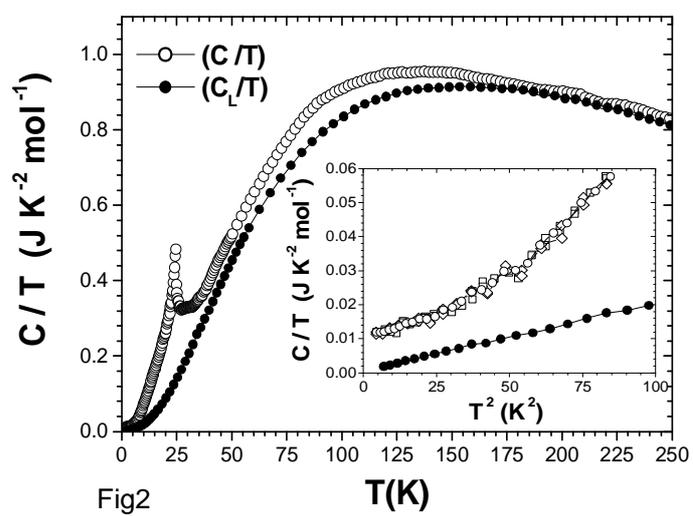
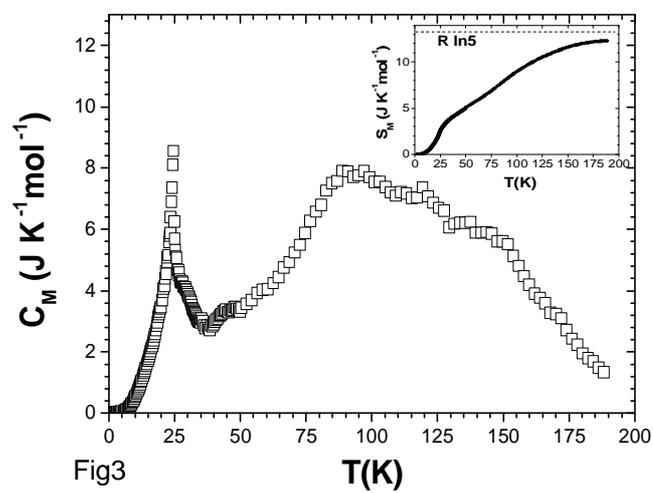
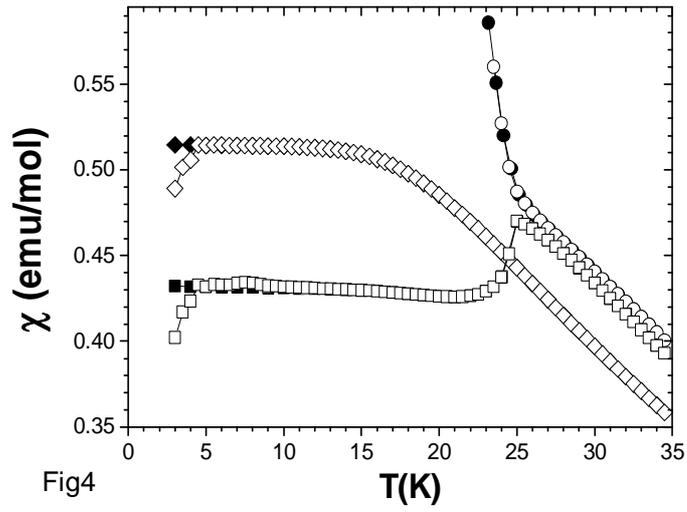
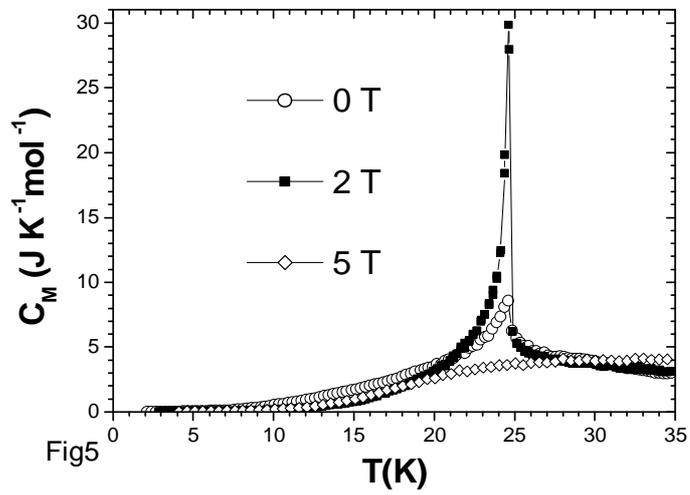
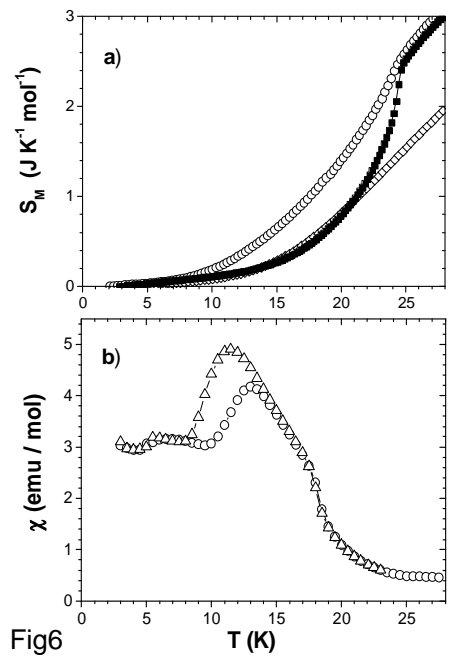
\end{document}